\documentclass{elsart}
\usepackage{t1enc}
\usepackage{amsmath}
\usepackage{epsfig}

\def\v#1{\underline{#1}}
\def\m#1{\mathbf{#1}}
\def\nelec{{N}} %\def\nelec{{N_{\text{electrons}}}}

\def\K{\mathcal{K}}

\def\comment#1{}
\def\eqref#1{Eq.\ \ref{eq:#1}}
\def\appref#1{Appendix\ \ref{app:#1}}
\def\tabref#1{Table\ \ref{tbl:#1}}
\def\secref#1{Sec.\ \ref{sec:#1}}
\def\unic{UNI\raisebox{1pt}{$\bullet$}C}

\newcommand{\rvec}{{\bf r}}
\newcommand{\rvecp}{{\bf r}'}
\newcommand{\rhor}{\rho ( \rvec )}
\newcommand{\vext}{v_{\text{ext}} ( \rvec )}
\newcommand{\veff}{v_{\text{eff}} ( \rvec )}

\newcommand{\psiri}{\psi_i ( \rvec )}
\newcommand{\psiriconj}{\bar{\psi}_i ( \rvec )}
\newcommand{\Hop}{\hat{H}}
\newcommand{\Sop}{\hat{S}}

\begin{document}
\begin{frontmatter}
\title{Solving large nonlinear generalized eigenvalue problems 
from Density Functional Theory calculations in parallel}

\author[unic]{Claus Bendtsen}
\author[camp]{Ole H.\ Nielsen} and
\author[camp]{Lars B.\ Hansen}
\address[unic]{UNI$\bullet$C, DTU Bldg.\ 304, DK-2800
Lyngby, Denmark} 
\address[camp]{Dept. of Physics, DTU
Bldg.\ 307, DK-2800 Lyngby, Denmark}

\begin{abstract}
The quantum mechanical ground state of electrons is described by
Density Functional Theory, which leads to large minimization problems.
An efficient minimization method uses a selfconsistent field (SCF)
solution of large eigenvalue problems.  The iterative Davidson
algorithm is often used, and we propose a new algorithm of this kind
which is well suited for the SCF method, since the accuracy of the
eigensolution is gradually improved along with the outer
SCF-iterations.  Best efficiency is obtained for small-block-size
iterations, and the algorithm is highly memory efficient.  The
implementation works well on both serial and parallel computers, and
good scalability of the algorithm is obtained.

\end{abstract}
\end{frontmatter}

\section{Introduction}

Within recent years it has become possible to perform quantum
mechanical calculations from first principles using the Density
Functional Theory (DFT) of Hohenberg, Kohn and Sham (see
e.g.\ \cite{paya89}).  Realistic calculations of the electrons'
ground-state can be carried out for large systems consisting of tens to
hundreds of atoms owing to improvements in physical methods, faster
computers, and more efficient algorithms.  In this way modeling
of a vast number of physical and chemical properties can be 
carried out, often taking large computer systems to their limits.

The Density Functional Theory specifies that the ground-state total
energy $E_0$ can be obtained as the global minimum of the energy
functional $E[ \rho ]$,
\begin{equation}
E[ \rho ] = T[ \rho ] + V_{\text{ee}} [ \rho ] + \int \rhor \vext d \rvec +
\gamma_{\text{Ewald}} \label{eq:dft}
\end{equation}
Here the position vector in space is denoted as $\rvec$, $\rhor$ (or
$\rho$ for short) is the charge density of electrons satisfying the
constraint $\int \rhor d \rvec = \nelec$, where $\nelec$ is the
integral number of electrons in the system.  The $\vext$ is the
``external'' potential function acting on the electrons and
originating from the atomic nuclei at fixed positions in the
system.  The kinetic energy $T[ \rho ]$ as well as the
electron-electron interaction energy $V_{\text{ee}} [ \rho ]$ are
functionals of the density $\rho$, only.  Finally, an electrostatic
repulsion energy between the nuclei, denoted as
$\gamma_{\text{Ewald}}$, is added.  The temperature is taken to be at
the absolute zero so that entropy terms are left out in the equations
below. 

Kohn and Sham\cite{paya89} proved that the charge density $\rhor$ can
be represented by a system of $\nelec$ {\em non-interacting}
electrons, whose complex-valued quantum-mechanical wavefunctions
$\psiri$ for $i=1,\ldots,\nelec$ yield the charge density as
\begin{equation}
\rhor = \sum_{i=1}^{\nelec} | \psiri |^2
\label{eq:rhor}
\end{equation}
(for simplicity, we only consider integral occupation numbers of 0 or 1)
and the non-interacting kinetic energy functional $T_s$ as
\begin{equation}
T_s [ \rho ] = - \half \sum_{i=1}^{\nelec} \int \psiriconj \nabla^2
\psiri d \rvec 
\end{equation}
($\psiriconj$ denotes the complex conjugate of $\psiri$).
The energy functional \eqref{dft} can now be rewritten as
\begin{equation}
E[ \rho ] = T_s[ \rho ] + E_{\text{Hartree}} [ \rho ] + E_{\text{xc}}
[ \rho ] + \int \rhor \vext d \rvec + \gamma_{\text{Ewald}}
\label{eq:Erho}
\end{equation}
where the $E_{\text{Hartree}} [ \rho ]$ is the classical electrostatic
energy of the electrons, and $E_{\text{xc}} [ \rho ]$ is an (unknown)
{\em exchange-correlation energy} functional.

The wavefunctions $\psiri$ are the $i=1,\ldots,\nelec$ lowest
eigensolutions of the {\em Kohn-Sham equation}
\begin{equation}
\Hop \psiri \equiv [ - \half \nabla^2 + \veff ] \psiri = \epsilon_i \psiri
\label{eq:kohnsham}
\end{equation}
where $\Hop$ is known as the Hamiltonian operator.  The {\em effective
potential} $\veff$ is the sum of the external, the {\em Hartree}
(electrostatic), and the {\em exchange-correlation} potentials:
\begin{equation}
\veff = \vext + 
\frac{\delta E_{\text{Hartree}}[ \rho ]}{\delta \rhor} +
\frac{\delta E_{\text{xc}}[ \rho ]}{\delta \rhor} =
\vext + \int \frac{\rho( \rvecp )}{| \rvec - \rvecp|} d \rvecp + 
v_{\text{xc}}( \rvec ) 
\label{eq:veff}
\end{equation}

It is seen from \eqref{veff} that the potential $\veff$ depends
on the charge density $\rhor$, which itself is given by
\eqref{rhor} as a sum of squared wavefunctions that are
determined as eigensolutions of \eqref{kohnsham}.  Hence a {\em
self-consistent} solution of these equations must be found starting
from an initial guess of $\rhor$, from which the $\veff$ is
constructed; solving the Kohn-Sham eigenvalue equation yields through
\eqref{rhor} a renewed charge density.  A minimization
algorithm for $E[ \rho ]$ must be used to obtain the ground-state
total energy $E_0$ from such iterations.

An additional complication in some of the current approaches
is the use of {\em ultrasoft} pseudopotentials\cite{va90} that puts a
lower requirement on the basis set size, but at the price of a more
complicated and computationally more demanding  Hamiltonian $\Hop$.
This leads 
to replacing the Kohn-Sham equation \eqref{kohnsham} by a {\em
generalized eigenvalue equation},
\begin{equation}
\Hop \psiri = \epsilon_i \Sop \psiri
\label{eq:ultrasoft}
\end{equation}
where the $\Sop$ denotes the overlap projection operator\cite{va90}.

In most numerical calculations the wavefunctions $\psiri$ are expanded
on a suitable, finite set of {\em basis functions}. The present work
employs the widely used {\em pseudopotential approximation} together
with {\em plane-wave} (or Fourier expansion) basis sets.
In the plane-wave basis the Hamiltonian \eqref{kohnsham} is diagonally
dominant owing to the kinetic energy term.
In this method full, complex Hermitian matrices of sizes
$10^3 - 10^5$ are usually
encountered, whereas the number of electrons $\nelec$ may typically be
two orders of magnitude smaller, $10^1 - 10^3$.
However, the algorithms described below remain valid for very general 
classes of problems using other basis function sets.

The translational symmetry of crystals is dealt with by 
introducing a summation over the so-called {\bf k}-points
(see e.g.\ \cite{krfu96}).  This leads to a set of independent
eigenvalue problems for each {\bf k}-point,
which are coupled only through the addition of
charge densities in \eqref{rhor}.  Similarly, electron spin
can double the number of independent eigenvalue problems
to be solved.
Such almost-decoupled eigenvalue problems should always be solved
independently in parallel, since this will be optimally efficient and can
easily be combined with the parallelization described below,
leading to the possibility of utilizing large numbers of processors
efficiently.
 
The discretized problem originating from \eqref{ultrasoft},
which essentially has to be solved for the smallest $\nelec$
eigenvalues, is the non-linear matrix eigenvalue problem,

\begin{align}
\m{H(\v{\rho})}\v{\psi_i} &= \epsilon_i \m{S} \v{\psi_i},\quad
i=1,\ldots,\nelec \label{eq:eig} \\
\v{\rho} &= \sum_{i=1}^{\nelec} \v{\psi_i}^H\m{S}\v{\psi_i},
\label{eq:rho}
\end{align}
where the overlap matrix $\m{S}$ is a symmetric positive definite
matrix when {\em ultrasoft} pseudo\-poten\-tials\cite{va90} are used
(and the identity matrix otherwise). This implies that the
eigenvalues, $\epsilon_i$ are real and the eigenvectors, $\v{\psi_i}$
mutually $\m{S}$-orthogonal.  

Given an initial charge density vector $\v{\rho}$ (reasonably good
initial values for $\rho$ can usually be constructed using the
densities of the constituent free atoms), \eqref{eig} can be solved
for the wave functions and the resulting charge density can
subsequently be computed using \eqref{rho}.  A new input charge
density can finally be formed, and the cycle can be iterated until
convergence. The overall procedure is commonly termed Self Consistent
Field (SCF).

Alternative approaches to the SCF method exist, notably direct 
minimization of the energy as a functional of the
wavefunctions $\psiri$ instead of the charge density $\rho$,
see for example \cite{tepaal89}.
It has been shown\cite{krfu96} that this method has about
the same efficiency as the traditional SCF method. 
Another active area of research tries to achieve order-N scaling by,
e.g., carefully selecting alternative basis functions.
However, a discussion of the numerical methods proposed in various
order-N methods are beyond the scope of this paper but a recent review
can be found in \cite{goed99}.

\section{Solving the Non-linear Eigenvalue Problem}

Using a quasi newton method for the iterations of \eqref{eig} and
\eqref{rho} generally results in convergence within a small number of
iterations.  The iterations are usually carried out by the method
Pulay\cite{pu80}, see e.g.\ \cite{krfu96} for a review of current methods. 
We propose a more robust and slightly more efficient approach, a quasi
newton method as described in \appref{secant}.  Experiments with the
limited memory BFGS method\cite{lino89} indicated that it has an
almost similar efficiency but is less robust.

Solving the eigenvalue problem \eqref{eig} is computationally by far
the most expensive part of the algorithm because of the cost of the
$\m{H(\v{\rho})}\v{\psi_i}$ product. The method of
Davidson\cite{davi75} has been especially successful for electronic
structure computations, and has been improved by iterating on several
eigenpairs simultaneously (see e.g.\ \cite{davi93}) and using better
preconditioners\cite{mosc86,oljosi90,sastchwuog96}. Lately, better theoretical
understanding of the methods has contributed to further
generalizations such as the Jacobi-Davidson method\cite{slvo96,sbfv96} and restarting techniques\cite{brpakn96,crphsa94,stsawu98}
along with other iterative methods for large eigenvalue problems based
on Rayleigh quotient iterations\cite{defagy98}, 
inverse iteration\cite{lalili97} or the Lanczos method\cite{mosc93,careso94}.
A currently advocated method in this field is the residual
minimization method, RMM-DIIS\cite{krfu96} but as this method essentially
just computes eigenpairs closest to the initial ones and therefore
easily can result in eigenpairs being missed, we cannot recommend this
approach. 

A major drawback---in the present context---of the
above methods is that they all (except for RMM-DIIS) focus on solving
the eigenvalue problem 
to a fairly high accuracy. However, this turns out not to be
necessary during every step of the selfconsistency cycle; in fact, our
calculations have shown that solving the eigenvalue problem to a high
accuracy generally does {\em not} decrease the number of overall steps
in the quasi newton/fixpoint iteration noticeably, and in certain cases even
increases the number of steps.

Thus we seek an eigenproblem solver which, given an initial
estimate of $\nelec$ eigenpairs, is able to improve this estimate (i.e.\
perform a relatively small number of iterations) for \textit{each} of
the eigenpairs.  In addition, an important aspect of large-scale
applications is that the memory requirement must be minimized.
In order to achieve these objectives, we investigated a number of algorithms
for improving eigenpairs so that residuals are
improved only until a certain, adjustable limit.

Using deflation techniques\cite{saad92,parl98} we experienced
convergence problems when deflating eigenpairs which were not well converged,
and therefore the overall eigenvalue problem had to be solved
to a much higher accuracy (and at a higher computational cost) than 
actually required.  Similar
experiences have been reported in \cite{sastchwuog96}, where the
convergence tolerance was loosened as more and more eigenpairs
converged. 

We instead propose an algorithm for solving the generalized
eigenvalue problem \eqref{eig}, which essentially is a blocked
Davidson-like algorithm using a non-orthogonal basis, but where a
maximum number of expansions on each eigenpair can be imposed. 

The reason for choosing a non-orthogonal basis, $\m{B}$ (and thus
obtaining a generalized eigenvalue problem in the projected space,
step \ref{projeig1} and \ref{projeig2}) is, besides from reducing the
computational cost compared to using an $\m{S}$-orthonormal basis,
that it reduces the number of synchronization points in the parallel
implementation (even when compared to an orthogonalization algorithm
implemented with delayed summation as in e.g.\
\cite{sastchwuog96}). The drawback is that projected eigenvalue
problem, step \ref{projeig1} and \ref{projeig2} could be poorly
conditioned and thus slow down the
convergence\cite[Sec. 11.10]{parl98}. Our initial basis (step
\ref{initialize}) is however well conditioned for every quasi newton
iteration (as the estimate of eigenvectors
$\v{\psi}_1,\ldots,\v{\psi}_\nelec$ is almost $\m{S}$-orthogonal and
$\m{S}$ is close to the identity) and as only a few expansions are
performed for each eigenpair (typically 2 to 3) a poorly conditioned
(projected) eigenvalue problem is unlikely to build up to the point
where it becomes a problem. This is due to the fact that the
eigenvectors do not have to be converged to a very high accuracy.
In fact, introducing orthogonalization in the algorithm has not led to
fewer iterations for a wide variety of tested physical problems.

The algorithm is as follows:

\begin{enumerate}
\item \label{initialize}
Given an estimate $\v{\psi}_1,\ldots,\v{\psi}_\nelec$ for the
$\nelec$ eigenvectors, initialize the subspace basis $\m{B} \equiv 
[\v{b}_1, \ldots, \v{b}_{\nelec}]=[\v{\psi}_1,\ldots,\v{\psi}_\nelec]$ 
and solve the small, projected subspace eigenvalue problem of dimension 
$\nelec$:
\[ 
\m{B}^H \m{H} \m{B}\v{\phi}_i=\lambda_i\m{B}^H \m{S} \m{B}\v{\phi}_i,
\quad i=1,\ldots,\nelec,
\text{ where }
\lambda_1\leq \lambda_2\leq \ldots \leq \lambda_\nelec. \label{projeig1}
\]
Select an iteration block-size $n_b$ and a maximum number of basis
vectors $n_{\text{max}}$ and a maximum number of iterations on each
eigenpair $k_{\text{max}}$.  \label{init}
\item Initialize the set $\K=\emptyset$. \label{outerit}
\item Loop over all eigenpairs, $i=1,\ldots,N$ 
\begin{enumerate}
\item If the $i$'th eigenpair $\lambda_i,\v{\phi}_i$ has less than
$k_{\text{max}}$ iterations then compute the corresponding residual
$\v{r}_i=\m{H}\m{B}\v{\phi}_i-\lambda_i\m{S}\m{B}\v{\phi}_i$.
If $\|\v{r}_i\|$ is larger than some tolerance,
add $i$ to $\K$. \label{tolerance} 
\label{residuals} 
\item If $\K$ has $n_b$ elements or $i=N$:
\begin{enumerate}
\item Do an iteration on eigenpairs given by indexes in $\K$:
\begin{enumerate}
\item ``Precondition'' the residuals\footnote{
The usual physical approximation in \cite{tepaal89} is used,
where the Hamiltonian is approximated by the diagonal
(and diagonally dominant) kinetic energy term in \eqref{kohnsham}.
A smooth function is multiplied onto the kinetic term in order to
damp the high-frequency errors.\cite{tepaal89}
}
corresponding to the unconverged eigenpairs,
i.e. solve approximately for
$\forall i\in \K:(\m{H}-\lambda_i\m{S})\v{x}_i=-\v{r}_i$.
\label{precond} 
\item Extend the basis, $\forall i\in\K:\m{B}=[\m{B},\v{x}_i/\|\v{x}_i\|]$. \label{lbl3}
\item Solve the updated subspace eigenproblem for the $\nelec$
smallest eigenvalues, \label{projeig2}
\begin{multline*}
\m{B}^H \m{H} \m{B}\v{\phi}_i=\lambda_i\m{B}^H \m{S} \m{B}\v{\phi}_i,
\quad i=1,\ldots,\nelec,\\
\text{ where }
\lambda_1\leq \ldots \leq \lambda_\nelec. 
\end{multline*}
\end{enumerate}
\item If the number of basis vectors in $\m{B}$ is too large
(i.e.\ larger than $n_{\text{max}}-n_b$) then collapse
the subspace (restart):
\[
\m{B}=\m{B}[\v{\phi}_1,\ldots,\v{\phi}_{\nelec}]. \]
\item Set $\K=\emptyset$.
\end{enumerate}
\end{enumerate}
\item Update
$[\v{\psi}_1,\ldots,\v{\psi}_\nelec]=\m{B}[\v{\phi}_1,\ldots,\v{\phi}_{\nelec}]$
\end{enumerate}

The tolerance of step \ref{tolerance} is set to an order of magnitude
smaller than the current residual norm of the quasi newton process
(i.e.\ $\|\v{r}^k\|_2$, page \pageref{resnorm}), but
no larger than $0.1$. The tolerance will therefore decrease with the
convergence of the quasi newton iteration.

Since the most time consuming part of the eigenvalue solver is computing
the products of $\m{H}$ and $\m{S}$ by vectors in the steps
\ref{projeig1}, \ref{residuals} and \ref{projeig2} these vectors are
stored and reused whenever possible.  Thus the dominant
memory requirement of
the algorithm is $\mathcal{O}(3n n_{\text{max}})$, where $n$ is the
vector length given by the dimension of \eqref{eig}.

Typical problem sizes might be $\nelec \approx 10^1 - 10^3$, and as it
will be shown in \secref{numexamp} the parameters should typically be
chosen so $k_{\text{max}}\approx 1 \text{ to } 2$, $n_b \approx 10$
and $n_{\text{max}} \approx N+5*n_b$, so the memory requirement is
very close to the absolute minimum, which is the $3 \nelec$ vectors
$\v{\psi_i}$, $\m{H}\v{\psi_i}$ and $\m{S}\v{\psi_i}$ for
$i=1,...,\nelec$.
  Lower memory
consumption could be achieved by eliminating the storage of
$\m{H}\v{\psi_i}$ and $\m{S}\v{\psi_i}$ and recomputing these products
as required in steps \ref{residuals} and \ref{projeig2}, however with
the tradeoff of requiring about twice as many multiplications of
$\m{H}$ and $\m{S}$ by vectors.

Most floating-point operations stem from the very complicated products
of $\m{H}$ and $\m{S}$ by vectors (the number of such operations
should be minimized by any algorithm).  In addition, the present
algorithm contains significant amounts of linear algebra
(approximately $8n\nelec$ flops for each $\m{H}$ product) which
however can be carried out efficiently using BLAS level 3 operations.

The asymptotic scaling of the total time spent on
the matrix-products for each quasi newton iteration is $\nelec n log ( n )$
due to the FFT-technique used, whereas the linear algebra
scales as $\nelec ^ 2 n$.  Since the matrix size $n$ increases roughly
linearly with $\nelec$, it is obvious that the linear algebra
would dominate asymptotically.  For the sizes of our problems, however,
the matrix products generally take longer than the linear algebra.

\section{Parallel Implementation}

The products of $\m{H}$ times vector parallelize particularly poorly,
since they involve 3D FFTs of fairly small size (typically $\approx
50$) in each dimension.  Therefore the parallelization for a
distributed memory computer is not viable solely across the plane
waves (vectors of length $n$).  By using a block size $n_b$ equal to
the number of available processors, it is however possible to perform
$n_b$ $\m{H}$-products in parallel using local FFTs exclusively --
each on a different processor (the $n_b$ may also be a multiple of the
number of processors, requiring additional memory, however).  This
approach is used in the implementation for the initial calculation of
the $\m{H}$ and $\m{S}$ products in step \ref{init} and for step
\ref{precond}, where the $n_b$ residual vectors each are gathered on
different processors before preconditioning them.  After the
preconditioning, $\m{H}$ and $\m{S}$ products are formed for the
preconditioned vector, and the results are finally redistributed
across the processors.  This requirement on the block size $n_b$ can
be lifted on computers that perform small-size parallel 3D FFTs with
high efficiency.

For the remaining part of the code the parallelization is performed
across the plane waves (i.e.\ each basis-vector is partitioned among
the processors).  This requires a fairly minimal amount of
communication (essentially only a small number of reduction
operations).

The parallel implementation was done using MPI\cite{mpi95}, and some
typical results indicating the scaling for runs carried at on the IBM
SP\footnote{Located at {\unic} in Lyngby, Denmark.} are given in
\tabref{parPt}\footnote{
This Platinum (Pt) problem is a an infinite slab with 35 atoms
in the unit cell.  The parameters used for the run were: vector
length $n=16409$ (25 Ry plane-wave cutoff energy),
$\nelec=210$, $n_b=\text{\# of processors}$,
$n_{\text{max}}=\nelec+8n_b$, number of selfconsistent iterations = 4.
The execution time is given for the eigensolver but with time for the
complete runs including initialization (not yet fully parallelized)
given in parenthesis.}.

\begin{table}[htbp] %\begin{center}
\caption{Parallel runs of the algorithm for a Pt problem. 
 \label{tbl:parPt} } 
\begin{tabular}{ccc} \hline
 \# of processors & execution time & speed-up \\ \hline
1  & 9544 (10932) & - \\ 
2  & 4208 (4671)& 2.3 \\
4  & 1860 (2192)& 5.1\\
6  & 1152 (1421) & 8.3 \\
8  & 901 (1197)& 10.6\\
12 & 634 (873)& 15.1 \\ \hline
\end{tabular} 
%\end{center} 
\end{table}

It is seen that the speed-up of the eigenvalue solver is very
satisfactory for these problem sizes, and larger problems are expected
to scale up to a correspondingly higher number of processors.  
Notice that the superlinear speed-up is partly due to the fact that
the blocksize and maximum subspace dimension varies with the number of
processors which implies that the 1 processor run did not benefit from
BLAS level 3 operations. A sequential run with the same parameters as
the 4 processor parallel run led to a speed-up of 3.0 on 4
processors. 

These runs were made using only a single ``{\bf k}-point''
(see Introduction).  In production runs the additional
trivial parallelism of a number of {\bf k}-points and spins
are utilized.  The parallelization of the eigenvalue solver 
is duplicated across mutually exclusive sets of processors 
for each {\bf k}-point and spin in an optimally efficient way.
Our work uses MPI communicators to implement this in a simple way.
However, memory consumption is not reduced by exploiting
this extra parallel dimension.

\section{Numerical Examples} \label{sec:numexamp}

\begin{figure}[htbp]
\begin{center}
\epsfig{figure=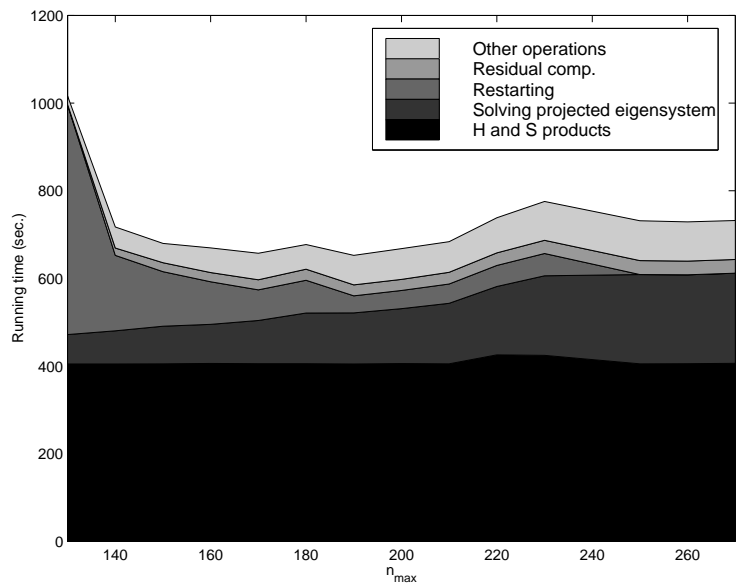,width=0.7\linewidth}
\caption{Time spent in different parts of the code for an Au problem,
using 6 processors ($n_b=6$), $k_{\text{max}}=1$ and varying the
maximum number of basis vectors, $n_{\text{max}}$.\label{fig:au_nmax}}
\end{center}
\end{figure}

\begin{figure}[htbp]
\begin{center}\epsfig{figure=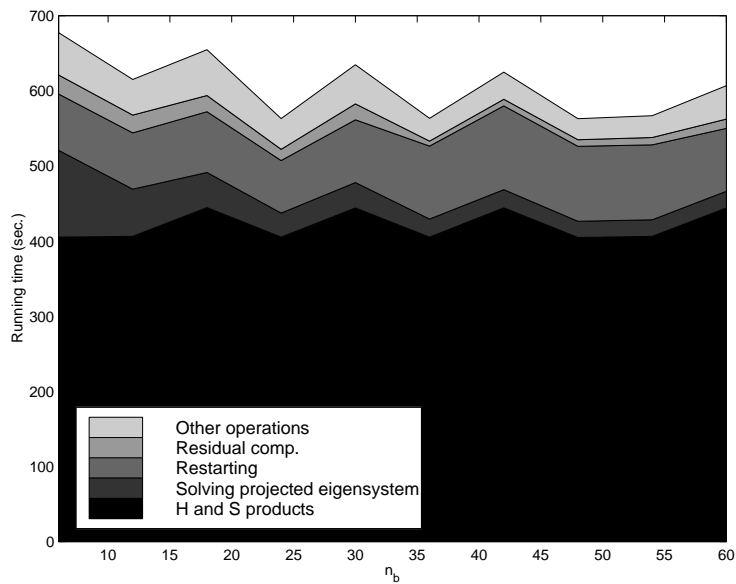,width=0.7\linewidth}
\caption{Time spent in different parts of the code for an Au problem,
using 6 processors, $k_{\text{max}}=1$, $n_{\text{max}}=180$ and varying the
blocksize, $n_b$.\label{fig:au_nb}}
\end{center}
\end{figure}

\begin{figure}[htbp]
\begin{center}
\epsfig{figure=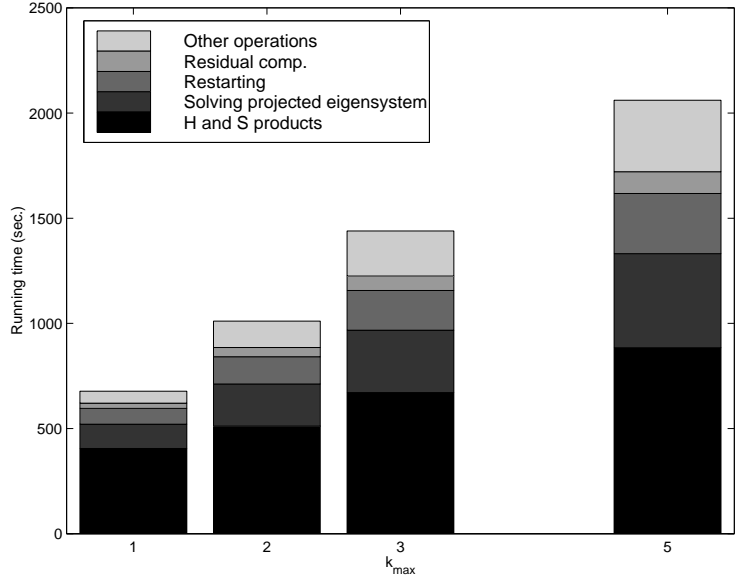,width=0.7\linewidth}
\caption{Time spent in different parts of the code for an Au problem,
using 6 processors ($n_b=6$), $n_{\text{max}}=180$ and varying the
maximum number of expansions on each eigenpair, $k_{\text{max}}$.\label{fig:au_kmax}}
\end{center}
\end{figure}

In order to investigate the stability of the proposed algorithm and
its sensitivity to choice of parameters, a large number of different
examples have been tested. Some representative examples\footnote{The
Gold (Au) problem shown in the figures
is an infinite slab with 18 atoms in the unit cell.
The parameters used for the run were: vector
length $n=13665$ (30 Ry plane-wave cutoff energy),
$\nelec=198$, number of selfconsistent iterations = 15.
} are shown in Fig.\ \ref{fig:au_nmax}, \ref{fig:au_nb} and
\ref{fig:au_kmax}.

The
experiments indicate that the size of $n_b$ is not critical as long as
it is chosen large enough for the solution of the projected
eigensystem not to be unnecessarily large. The maximum subspace
dimension, $n_{\text{max}}$ is however important -- if
$n_{\text{max}}$ is too small the overhead in restarting becomes
dominant, on the other hand if $n_{\text{max}}$ becomes unnecessarily
large the memory consumption and the cost of solving the projected
eigenvalue problem becomes large.

\begin{figure}[htbp]
\begin{center}
\epsfig{figure=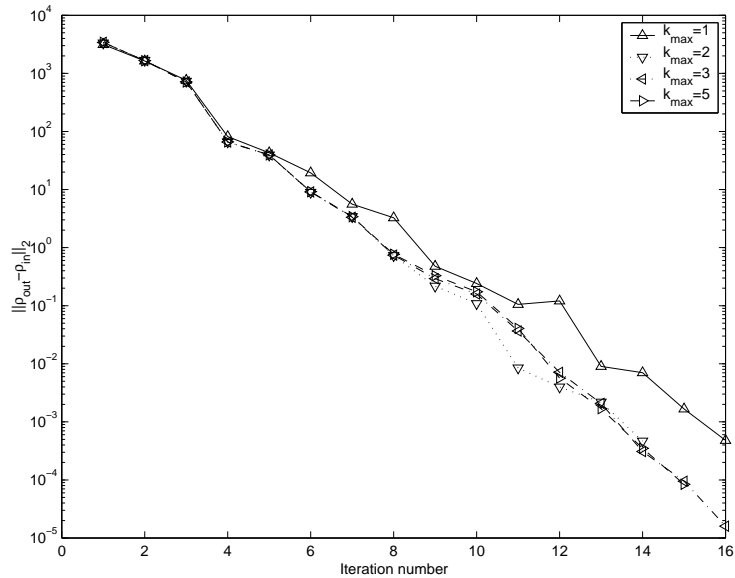,width=0.7\linewidth}
\caption{Convergence of the quasi newton process for an Au problem for
varying maximum number of expansions on each eigenpair,
$k_{\text{max}}$.\label{fig:convergence}}
\end{center}
\end{figure}

From Fig.\ \ref{fig:au_kmax} it is seen that the overall work grows
linearly when increasing the precision to which we solve the
eigenvalue problem, indicating that it is not worthwhile to solve the
eigenvalue problem to a high accuracy. This is furthermore supported
by Fig.\ \ref{fig:convergence} which shows that the convergence of the
quasi newton process is largely unaffected by increasing the number of
iterations on each eigenpair, $k_{\text{max}}$.

\section{Conclusion}

The algorithm proposed in this paper presents an efficient way to solve the
nonlinear generalized eigenvalue problems occuring in DFT calculations
on multiple processors. Part of the algorithm is a variant of the
Davidson algorithm\cite{davi93} which focuses on using only small
amounts of work for each eigenpair in each of the quasi newton
iterations, and thereby being able to use a non-orthogonal basis. 

In contrast, our best effort using traditional deflation techniques
typically required 3 to 5 times the amount of overall work, as each of
the eigenpairs had to be converged to a much higher precision in order
for the deflation to be robust. Attempts were made to improve on this
and an algorithm using deflation which was almost as efficient as the
algorithm proposed on page \pageref{initialize} was devised.  However,
it involved complicated heuristics especially for treating eigenvalues
with higher multiplicity (which often occur due to symmetries in the
systems) and this additional effort did not seem worthwhile.

In summary, the algorithm proposed in the present paper has been
designed for nonlinear eigenvalue problems such as those occurring in
selfconsistent DFT calculations.  The algorithm has shown robustness
and efficiency, is flexible in accomodating memory limitations, and
allows constraints in the number of iterations and the tolerance of
eigenpairs.  Several parameters can be used to optimize the
convergence, and we propose recommended values for DFT iterations.
Good parallel performance is achieved on distributed memory computers,
but since the eigenvalue problem in the projected subspace is not
parallelized, the parallel scalability may be limited, independent of
the efficiency of the interconnecting network.

%\section*{Appendix}

\appendix

\section{The Projected Newton Method} \label{app:secant}

Given a series of input charge densities,
$\v{\rho}_{\text{in}}^{k-m},\ldots,\v{\rho}_{\text{in}}^k$  
($k \geq m \geq 1$) for \eqref{eig}, the corresponding output charge
densities defined by \eqref{rho} will be denoted as
$\v{\rho}_{\text{out}}^{k-m},\ldots,\v{\rho}_{\text{out}}^k$. The residual
of the charge density iteration is then defined by
\[
\v{r}^k = \v{\rho}_{\text{out}}^k - \v{\rho}_{\text{in}}^k \label{resnorm}
\]
of which the Jacobian can be approximated by
$\m{T}_k\m{G}_k^{\dagger}$, where $\m{G}^{\dagger}$ denotes a
pseudo inverse of $\m{G}$, and we have defined
\begin{align*}
\m{G}_k &= (\v{\Delta\rho}^{k-m+1},\ldots,\v{\Delta\rho}^k), & 
\v{\Delta\rho}^k &= \v{\rho}_{\text{in}}^k -
\v{\rho}_{\text{in}}^{k-1} \\
\m{T}_k &= (\v{\Delta r}^{k-m+1},\ldots,\v{\Delta r}^k), & 
\v{\Delta r}^k &= \v{r}^k - \v{r}^{k-1}.
\end{align*}
Applying Newton's method gives 
\begin{equation}
\v{\rho}_{\text{in}}^{k+1} = \v{\rho}_{\text{in}}^k -
\m{G}_k\m{T}_k^{\dagger}\v{r}^k \label{eq:newton}
\end{equation}
where $\v{\alpha}_k\equiv \m{T}_k^{\dagger}\v{r}^k$ is obtained as the
least squares solution of 
\begin{equation}
\m{T}_k \v{\alpha}_k = \v{r}^k. \label{eq:lsq}
\end{equation}

It is seen that the fixpoint solution $\v{\rho}^*\equiv \v{\rho}^k$ corresponding
to $\v{r^k}=\v{0}$ will never be found unless
$\v{\rho}^*-\v{\rho}^i$ (where $i$ is equal to the initial $k$ used)
belongs to the span of $\m{G}_i$ 
(i.e., the span of $\m{G}_k$ remains constant).
Therefore, a term
$\m{\beta}(\v{r}^k-\m{T}_k\v{\alpha}_k)$ is added to the right hand
side of \eqref{newton} for some arbitrarily chosen full rank, 
diagonal matrix $\m{\beta}$
as proposed in \cite{an65} (notice that $\v{r}^k=\m{T}_k\v{\alpha}_k$
if $\m{T}_k$ has full rank).

The least squares system \eqref{lsq} is solved via the thin QR
factorization on the augmented matrix $[\m{T}_k \v{\alpha}_k]$ as
described in e.g.\ \cite{bj96}, and the factorization is updated rather than
recomputed in the following iteration using the routines from \cite{regr90}.

As the above algorithm can not be used for the first iteration, 
the value $\v{\rho}_{\text{in}}^{1} =
\v{\rho}_{\text{in}}^0+\m{\beta}\v{r}^0$ is used instead. For the subsequent
iterations the above procedure is applied with $m$ incremented by one
for each iteration (starting with $m=1$) until a maximum value
$m_{\text{max}}$ is reached. Choosing $m_{\text{max}}$ in the range 
of $5$ to $10$
generally works well for the present DFT problems.

\bibliography{../main}
\bibliographystyle{amsplain}

\end{document}